\documentstyle[multicol,prl,aps,epsfig]{revtex}
\begin{document}
\draft
\title{\bf Non-Gaussianity of Resistance Fluctuations 
\\ Near Electrical Breakdown}

\author{C. Pennetta$^{1,2}$, E. Alfinito$^{1}$, 
L. Reggiani$^{1,2}$ and S. Ruffo$^{3}$}

\address{
$^{1}$ INFM - National Nanotechnology Laboratory, Via Arnesano, 73100, Italy.
\\ $^2$ Dipartimento di Ingegneria dell'Innovazione, Universit\`a di Lecce,  
Via Arnesano, 73100, Italy.
\\ $^3$ INFN, INFM and Dipartimento di Energetica ``Sergio Stecco'', 
Universit\`a di Firenze, \\Via S. Marta, 3, Firenze, 50139, Italy.
}
\date{\today}
\maketitle

\begin{abstract}

We study the resistance fluctuation distribution of a thin film near 
electrical breakdown. The film is modeled as a stationary resistor network 
under biased percolation. Depending on the value of the external current, 
on the system sizes and on the level of internal disorder, the fluctuation 
distribution can exhibit a non-Gaussian behavior. We analyze this 
non-Gaussianity in terms of the generalized Gumbel distribution recently 
introduced in the context of highly correlated systems near criticality. 
We find that when the average  fraction of defects approaches the random 
percolation threshold, the resistance fluctuation distribution is well 
described by the universal behavior of the Bramwell-Holdsworth-Pinton 
distribution. 

\end{abstract}
\begin{multicols} {2}

\section{Introduction and Model}
Electrical breakdown is an irreversible change in the resistance of a two 
terminal device and it usually occurs in the presence of high applied electric
stress, like current or voltage. Breakdown of condunctor-insulator composites 
\cite{bardhan}, soft dielectric breakdown in ultrathin oxide films 
\cite{ohring} and degradation of metallic films due to electromigration 
\cite{ohring,pen_em} are typical experimental examples. Nonlinear current 
voltage characteristics, giant enhancement of excess noise over the quadratic 
dependence on the applied field are the mostly studied breakdown precursors. 
\cite{bardhan,pre_fnl}. Yet another important and less understood breakdown 
precursor is represented by the emergence of non-Gaussian fluctuations 
\cite{weissman}, which origin has attracted an increasing interest in the 
recent literature \cite{bramwell_nat,bramwell_prl,bramwell_pre,z_racz}. 
For these reasons, here we study the distribution of the resistance 
fluctuations of a thin film near electrical breakdown. To this purpose 
we make use of the Stationary Network Under Biased Percolation (SNUBP) 
model \cite{pen_em,pre_fnl}. This model allows the study of the electrical 
conduction of disordered materials over the full range of applied bias values,
from the linear regime up to the breakdown. Moreover, the SNUBP provides a 
good modeling of many features associated with the electrical instability of 
composites materials \cite{bardhan} and with the electromigration damage of 
metal lines \cite{pen_em}, two important classes of breakdown phenomena.

In the SNUBP model \cite{pen_em,pre_fnl}, the resistance and the resistance 
fluctuations of a disordered film are determined by the competition of two 
biased processes taking place in a two-dimensional square-lattice resistor 
network. More precisely, the network consists of $2N^2$ resistors in two 
possible states: (i) regular, corresponding to resistors with resistance 
$r_n =r_0[1+\alpha(T_n - T_0)]$ and (ii) broken, corresponding to resistors 
with $r_{OP} = 10^9 r_n$. Here, $N$ determines the linear sizes of the network,
$\alpha$ is the temperature resistance coefficient, $T_0$ the bath 
temperature, and $T_n$ the local temperature of the n-resistor, resulting from
Joule heating and thermal exchange with neighbour resistors \cite{prl_fail}:
\vspace*{-0.4cm}
\begin{equation}
T_{n}=T_{0} + A \Bigl[ r_{n} i_{n}^{2} + {3 \over 4N_{neig}}
\sum_{l=1}^{N_{neig}}  \Bigl( r_{l} i_{l}^2   - r_n i_n^2 \Bigr) \Bigr]
\label{eq:temp}
\end{equation}
%
where $N_{neig}$ is the number of first neighbors around the n{\em th} 
resistor and $i_{n}$ the current flowing in it. The parameter $A$ represents 
the thermal resistance of each resistor and it determines the importance of 
Joule heating effects. The network is contacted at the left and right hand 
sides to perfectly conducting bars through which a constant stress current 
$I$ is applied. The two biased processes consist of stochastic transitions 
between the two possible states and they occur with thermally activated 
probability, $W_D$ and $W_R$, which depend on the local temperature and are 
characterized by two activation energies, $E_D$ and $E_R$, with $E_D > E_R$ 
for networks of finite sizes \cite{pre_fnl,prl_stat}. The network evolution 
is obtained by a Monte Carlo simulation which updates the network resistance 
after breaking and recovery processes according to an iterative procedure as 
detailed in Ref. \cite{pre_fnl}. The sequence of the successive configurations
provides the final $R(t)$ signal with the time scale calibrated on the 
iteration steps. Then, depending on the $E_D$ and $E_R$ values and on the 
stress conditions, the network reaches a steady state or undergoes an 
irreversible electrical failure. This last possibility is associated with the 
existence of at least one continuous path of broken resistors between the 
upper and lower sides of the network, i.e. with the achievement of the 
percolation threshold, $p_C$, for the fraction of broken resistors. 
It has been found \cite{pre_fnl} that for a given value of 
$T_0$, a threshold current value, $I_B$, exists above which the electrical 
breakdown occurs. On the other hand, for $I \le I_B$ the network is stable, 
i.e. the fraction of broken resistors, $p$, and the network resistance, $R$, 
fluctuate around their average values $<p>$ and $<R>$, respectively. In the 
following section we will analyze the results of simulations carried out by 
considering square  networks of size $75 \times 75$ and by taking the 
following values for the other parameters (chosen as physically plausible): 
$r_0=1$ ($\Omega$), $\alpha = 10^{-3}$ (K$^{-1}$), $A=5 \times 10^5$ (K/W), 
$E_D = 0.170$ (eV), $E_R$ in the range $0.026 \div 0.155$ (eV) and $T_0=300$ 
(K). The values of the external current range from $0.001 \le I \le 3.0$ (A). 
\vspace*{-0.6cm}
\section{Results}
Figure 1 reports a set of resistance evolutions for increasing current values.
In this case the activation energy for the recovery process is $E_R = 0.043$
(eV). This value of $E_R$ implies a breakdown current $I_B=2.1$ (A). The 
figure displays two important features of the electrical response of 
conducting films to high fields: i) the average resistance exhibits an Omic 
behavior at low currents and then increases significantly at high currents 
(non-Ohmic regime); ii) the amplitude of resistance fluctuations increases 
strongly for currents near to the breakdown value. By analyzing the first and 
the second moment of the steady resistance we have found that the non ohmic 
regime can be divided in two regions \cite{pre_fnl}. Actually, a moderate bias
region, where both the average resistance $<R>$ and the relative variance of 
resistance fluctuations $\Sigma \equiv<(\Delta R)^2>/<R>^2$ increase 
quadratically with the current, is followed by a pre-breakdown region where 
superquadratic behaviors emerge \cite{pre_fnl}.  
 \begin{figure}[bth]
 \vspace*{-0.2cm}
 \epsfig{figure=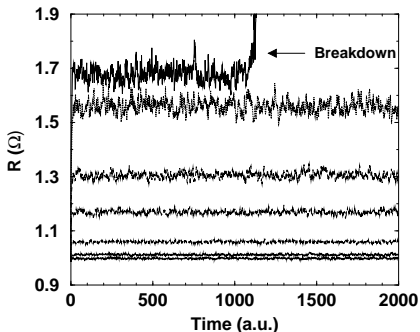,width=5.5cm}
 \vspace*{-0.1cm}
 \caption{Resistance evolutions for increasing bias values. Starting from
 the bottom, the six curves correspond to steady states at 
 $I=0.01, 0.5, 1.0, 1.5, 1.8, 2.1$ (A). The highest curve 
 corresponds to breakdown and it is obtained for $I=2.2$ (A).}
 \label{fig:1}
\end{figure}
Remarkably, another feature of the resistance fluctuation
distribution is found to emerge: the non-Gaussian behavior which accounts for
the importance of moments higher than the second. Here, we have thus 
investigated the Gaussianity of the $R(t)$ steady-state signals shown in 
Fig. 1. We have considered time series containing about $3 \times 10^5$ 
resistance values. The results of the analysis are reported in Fig. 2 which 
shows the probability density function (PDF), $\phi$, of the distribution of
resistance fluctuations for two values of the bias current: $I=0.01$ (A)
(triangles) and $I=I_B=2.1$ (A) (full circles). Precisely, in this figure
we have reported on a lin-log plot the product $\sigma \Phi$ as a function of
$(<R>-R)/\sigma$ (where $\sigma$ is the root mean square deviation). 
In fact, by making the distribution independent of its first and second 
moments, this normalized representation is particularly convenient to  
explore the functional form \cite{bramwell_nat}. 
 \vspace*{-0.2cm}
 \begin{figure}[bth]
 \epsfig{figure=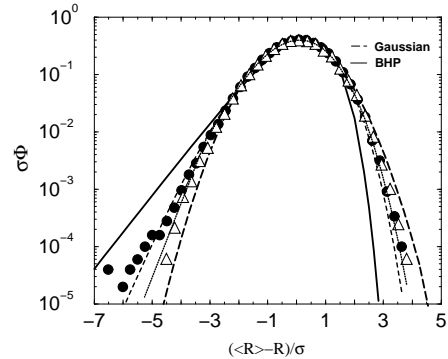,width=5.8cm}
 \caption{Scaled PDF of resistance fluctuations,
 $\Pi=\sigma \phi$, for two current values: $I=0.01$ (A) (triangles) and 
 $I=2.1$ (A) (full circles). The other curves are respectively: BHP
 distribution (thick solid line), Gaussian (long dashed), best-fit to the
 $I=0.01$ (A) data (solid), best-fit to the  $I=2.1$ (A) data (dashed).}
 \label{fig:2}
 \end{figure}
For comparison, in Fig. 2 we also show the Gaussian distribution (long dashed 
line), which in this normalized representation has zero mean and unit 
variance. We notice that in 1998  Bramwell, Holdsworth and Pinton 
\cite{bramwell_nat} realized that the PDF of a global quantity of a system at 
a critical point takes a universal behavior, irrespectively
of the particular quantity and of the nature and sizes of the system. 
Successively, the functional form describing this universal behavior was 
identified \cite{bramwell_prl}. Therefore, in Fig. 2 we report also the 
Bramwell, Holdsworth and Pinton (BHP) distribution (thick continuos line). 
Before giving its expression, we have to introduce the following definitions 
\cite{bramwell_prl}. We call $m$ a fluctuating quantity (for example the 
magnetization of a ferromagnet like in Ref. \onlinecite{bramwell_prl}), $<m>$ 
and $\sigma_m$ its mean value and root mean square deviation respectively, 
$P(m)$ its PDF, $y\equiv (m-<m>)/\sigma_m$ the normalized variable, 
$\Pi(y)\equiv \sigma_m P(y)$ the normalized PDF and finally $x \equiv b(y-s)$.
Then, the BHP distribution takes the following expression \cite{bramwell_prl}:

\vspace*{-0.2cm}
\begin{equation}
\Pi(y) = K [e^{x - e^{x}}]^a \label{eq:bhp}
\end{equation} 
where $a=\pi/2$, $b=0.936 \pm 0.002$, $s=0.374 \pm 0.001$ and $K=2.15 \pm0.01$.
Equation ~(\ref{eq:bhp}) can be considered a generalization of the Gumbel
distribution, which is often associated with the occurrence of rare events. 
Thus, looking at Fig. 2 we can outline the following main points: 
i) the resistance fluctuation distribution exhibits a non-Gaussian behavior; 
ii) the non-Gaussianity increases for currents near to the 
breakdown value; iii) even at low biases (i.e within the Ohmic regime) 
the distribution keeps a weak  non-Gaussianity; iv) the distribution at $I_B$ 
departs significantly from  the universal behavior. Point iii) is understood 
as a finite size effect, as confirmed by the analysis of data for networks of
different sizes,  not reported here. Thus, the points iii) and iv), by 
indicating a size dependent and non universal behavior, both witness a first 
order transition, in agreement with the results in Refs. 
\onlinecite{pre_fnl,upon02} and with electrical breakdown experiments in 
composites \cite{bardhan}. On the other hand, by recalling that real systems 
have finite sizes, the feature ii) can provide an important precursor of 
failure for systems of finite and given sizes. In order to identify what kind 
of distribution describes the data in Fig. 2, we have considered several 
distributions among those available in the literature. Accordingly, we have 
found that the best description to the present data is given by the expression
in Eq.~(\ref{eq:bhp}) once the parameters $a$, $b$, $s$ and $K$ are considered
as fitting parameters. The dashed and the solid curves in Fig. 2 represent the 
best-fit with Eq.~(\ref{eq:bhp}) to the data for $I=2.1$ (A) (full-circles) 
and for $I=0.01$ (A) (triangles), respectively.

\vspace*{-0.2cm}
 \begin{figure}[bth]
 \epsfig{figure=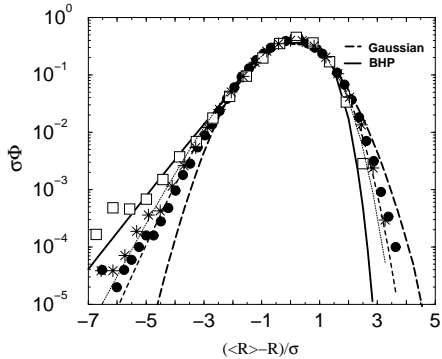,width=5.8cm}
 \caption{Scaled PDF of resistance
 fluctuations, $\Pi=\sigma \phi$, for different values of the recovery
 energy. The full circles correspond to $E_R=0.043$ (eV), the stars to
 $E_R=0.130$ (eV) and the squares to  $E_R=0.155$ (eV). In all cases 
 $I=I_B$. The other curves are respectively: BHP distribution (thick solid
 line), Gaussian (long dashed), best-fit to the $E_R=0.043$ (eV) data
 (dashed), best-fit to the $E_R=0.155$ (eV) data (solid).}
 \label{fig:4}
 \end{figure}
 \vspace*{-0.2cm}

In the following, we further consider the effect of the activation energy 
of the defect recovery process on the PDF. In fact, the recovery energy, by 
controlling the level of disorder inside the system, plays an important role
on the system behavior \cite{pre_fnl}. To this purpose, Fig. 3 reports the 
distributions of the resistance fluctuations obtained for different values of 
the recovery energy: $E_R=0.043$ (eV) (full circles), $E_R=0.130$ (eV) 
(stars), $E_R=0.155$
 (eV) squares. In all the cases, the current corresponds
to the breakdown, and the values are: $I=2.1$ (A), $I=0.38$ (A) and $I=0.095$ 
(A), respectively. Thus, full circles are the same data of Fig. 2.
We notice that $E_R=0.155$ (eV) is very close to the $E_D$ value and 
practically represents one the largest value of $E_R$ yet providing 
a steady state of the network \cite{pre_fnl}. Therefore, as discussed in
Ref. \cite{pre_fnl}, for this value of $E_R$ the network is very near to its 
critical point and it is characterized by an average fraction of defects 
($<p>=0.35$) close to the value of the random percolation threshold (0.5 for 
bond percolation on a square and infinite lattice). Accordingly, from Fig. 3 
we can see that the resistance fluctuation distribution corresponding to this 
value of $E_R$ is well described by the universal BHP distribution. The solid 
and the dashed curves in this figure represent the best-fit with 
Eq.~(\ref{eq:bhp}) to the data obtained for $E_R=0.130$ (eV) and $E_R=0.043$ 
(eV), respectively. Also in these cases the fit with Eq.~(\ref{eq:bhp}) is 
found to be satisfactory.

%
In conclusions, we have studied the distribution of resistance fluctuations
of a thin film under different bias conditions and for several
values of the recovery energy. The distribution exhibits a non-Gaussian
behavior which is well described by a generalized form of the Gumbel 
distribution. Furthermore, a value of the recovery energy close to that of 
defect generation, a condition which corresponds to an average defect 
fraction approaching the random percolation value, is identified as the 
condition under which the distribution achieves the universal behavior 
described by the BHP distribution \cite{bramwell_nat}.

\vspace*{-0.5cm}
\section*{ACKNOWLEDGEMENTS}
\vspace*{-0.5cm}
This research has been  performed within the SPOT-NOSED project IST-2001-38899
of the EC.

\vspace*{-0.7cm}

\end{multicols}

\newpage\noindent
%
%
%

\end{document}